\documentclass[twocolumn,
longbibliography,
showpacs,                     %  Add 'showpacs' option to make PACS codes appear
showkeys,                  %  Add 'showkeys' option to make keywords appear
secnumarabic,
amssymb, 
nobibnotes, 
floatfix,
aps, 
pre]{revtex4-1}

\usepackage{graphicx}
\usepackage{latexsym}
\usepackage{amsmath} 
\usepackage{verbatim}

\newcommand{\pd}[2]{\frac{\partial #1}{\partial #2}}

\newcommand{\pdbs}[3]{\left(\frac{\partial #1}{\partial #2}\right)_{#3}}

\begin{document}

\title{Resolving the debate about proposed expressions for the classical entropy}

\author{Robert H. Swendsen}
\email[]{swendsen@cmu.edu}
\affiliation{Department of Physics, Carnegie Mellon University, Pittsburgh PA, 15213, USA}
%\thanks{}
%\altaffiliation{}

%\pacs{05.70.-a, 05.20.-y}% PACS, the Physics and Astronomy
                             % Classification Scheme.
\keywords{Entropy; Boltzmann; Gibbs; canonical; grand canonical; classical ideal gas}%Use showkeys class option if keyword
                              %display desired
                              
\date{\today}

\begin{abstract}
Despite well over a century of effort,
 the proper expression 
for the classical entropy
 in statistical mechanics 
remains a subject of debate.
The Boltzmann entropy
(calculated from a surface in phase space)
has been criticized as  not being an adiabatic invariant.
It has been suggested 
that the Gibbs entropy (volume in phase space)  is correct,
which would forbid the concept of 
 negative temperatures.
An apparently innocuous assumption 
turns out to be  responsible for much of the controversy,
namely,
 that the energy $E$ and the number of particles $N$
are given exactly. 
The true distributions are  known 
to be extremely narrow
(of order $1/\sqrt{N}$),
so that
it is surprising that this is a problem.
The 
 canonical and grand canonical ensembles
provide
  alternative expressions 
 for the entropy  
 that satisfy all requirements.
The consequences are that 
negative temperatures are thermodynamically valid,
the validity of the Gibbs entropy is limited to increasing densities of states,
and the completely correct expression for the entropy
is given by the grand canonical formulation.
The Boltzmann entropy
is shown to 
provide an excellent approximation 
in almost all cases.
\end{abstract}

\maketitle

\section{Introduction}
\label{introduction}

Entropy is such a fundamental concept in the fields of statistical
mechanics and thermodynamics that one might think its definition could not be a matter of dispute.
Nonetheless, the proper definition of entropy has become a matter of heated discussion.
Some workers in the field
have pointed to weaknesses in the Boltzmann definition
(defined in section \ref{section_Boltzmann})\cite{Boltzmann,Boltzmann_translation},
and claimed that the Gibbs %(volume) 
entropy 
(defined in section \ref{section_Gibbs})\cite{Gibbs,Hertz_1910,Gibbs_entropy_2}
must replace it\cite{Berdichevsky_et_al_1991,%  Negative temperature of vortex motion
Campisi_SHPMP_2005,%
DH_Physica_A_2006,%   {Phase transitions in small systems: Microcanonical vs. canonical ensembles
Campisi_Kobe_2010,%
Romero-Rochin_2013,%
Sokolov_2014,%
DH_NatPhys_2014,%  Consistent thermostatistics forbids negative absolute temperatures
DH_reply_to_FW,%
HHD_2014,%   S. Hilbert and P. H\"anggi and J. Dunkel
Campisi_SHPMP_2015,%  Construction of microcanonical entropy on thermodynamic pillars
HHD_2016%   P. H\"anggi and  S. Hilbert and J. Dunkel
}.
Others have defended the use of the Boltzmann entropy
and pointed to flaws in the Gibbs entropy\cite{Miranda_2015,%
Frenkel_Warren_2015,%
Vilar_Rubi_2014,%
Schneider_et_al,%
DV_Anghel_2015,%
Cerino_2015,%
Poulter_2015,%
SW_2015_PR_E_R,%
Wang_2015,%
SW_2016_Physica,%
RHS_continuous,%
Matty_comparison,%
Buonsante_et_al_2016,%  Buonsante, P., Franzosi, R., Smerzi, A.
Buonsante_et_al_2017,%  Buonsante, P., Franzosi, R., Smerzi, A.
RHS_entropy_def_2017,
Abraham_Penrose_2017%
}.

At the center of the debate is the question 
of whether negative temperature\cite{Purcell_Pound,Ramsey_Neg_T}
is a thermodynamically consistent concept.
The Gibbs entropy 
requires the temperature 
to be positive\cite{Campisi_SHPMP_2005,%
DH_Physica_A_2006%   {Phase transitions in small systems: Microcanonical vs. canonical ensembles
},
while the Boltzmann entropy allows it to be negative.

Although most of the discussion has been restricted 
to the Boltzmann and Gibbs entropies,
I believe that this restriction is responsible 
for the controversy.
%
%weaknesses in both the Boltzmann and Gibbs entropies.
Since the two forms of entropy have different strengths and weaknesses,
the two sides have disagreed mainly on 
which strengths and weaknesses
are most important.
The argument can  be resolved by 
opening up the discussion
to consideration of other  definitions of entropy.
In this paper,
I will examine the strengths and weaknesses   of    
the Boltzmann and Gibbs entropies,
and discuss two alternatives
%that are superior to both.
that do not have the weaknesses of either.
By freeing the discussion from an artificial restriction 
in the definition,
we will see that thermodynamics
can provide a thermodynamically consistent 
basis for negative temperature

The term ``entropy'' has been associated with many related,
 but different concepts.
 I am concerned exclusively with the thermodynamic entropy.
I  will limit the present discussion to classical systems.
Although many of the disagreements 
concern quantum systems,
they are largely based on classical arguments.

The discussion is also limited to systems that are finite,
but not small.
Although claims have been made that 
thermodynamics should apply to small systems
-- even a system consisting of a single particle\cite{HHD_2014,HHD_2016} --
I will  only consider  macroscopic systems,
in which there are sufficient particles 
for the fluctuations to be  smaller than 
the resolution of experiments,
as a rough estimate, more than $10^{12}$ particles. 
For arguments to be valid for small systems,
they must at least be valid
for large, finite systems.

%A central result is that 
The main source of disagreement is %due to 
the fact 
that the microcanonical ensemble 
is not quite correct,
although the discrepancies in most cases are very small.
The errors in the Boltzmann and Gibbs entropies 
%for most systems
are usually of the order of $1/N$
or $\ln (N) /N$.
Such errors are masked by fluctuations 
of order $1/\sqrt{N}$,
and, indeed,
are not  measurable in a 
macroscopic system\cite{SW_2016_Physica,unmeasurable_error}.
These errors must nevertheless be taken seriously.
The advocates 
of the Gibbs entropy have 
argued
%based their claims
%on arguments 
 that 
the Boltzmann entropy violates thermodynamics
for finite, classical systems
because of
%For systems with a density of states that decreases  with energy  the consequences become very important.
discrepancies
%with exact results 
of the order of $1/N$  in the predicted energies 
%have been claimed to justify the abandonment 
%of the Boltzmann entropy
\cite{
Campisi_SHPMP_2005,%
DH_Physica_A_2006}.
I will therefore examine the exact behavior of classical systems,
even if the errors are as small as order $1/N$.

%%%%%%%%%%%%%%

The practical purpose of thermodynamics 
is to  predict 
 the results of experiments.
These predictions are ultimately based on probabilities 
calculated in statistical mechanics. 
For example, 
the probability distribution of 
the energy could be calculated by statistical mechanics
 for a system had previously been in contact with another system.
This calculated distribution could,
in principle,
be checked experimentally by many repetitions of the experiment,
bringing two systems into thermal contact and then separating them.
This is the concept of probability that I will use throughout.

Boltzmann was the first to establish a connection 
between thermodynamics    
%experiment
and the probability of observing values
of extensive variables
in experiments%
%in equilibrium
\cite{Boltzmann,%
Boltzmann_translation,%
Boltzmann_classical_entropy,%
RHS_4}.
The derivation 
of the probability distribution  
in Section \ref{Macroscopic_probabilities}
and the corresponding expression for the Boltzmann entropy
in Section \ref{section_Boltzmann}
is a generalization of 
Boltzmann's idea 
to include the volume and the particle number as 
thermodynamic variables.

%%%%%%%%%%%%%%%

There is a  error 
due to the use of the  microcanonical ensemble:
It
gives expressions for the entropy  
that violate an exact  thermodynamic stability condition
% for first-order  phase transitions
 \cite{Matty_comparison,%
Griffin_2017%
}.
I will show that this is also corrected 
by the canonical and grand canonical entropies
in section \ref{first_order_canonical}.

In this paper,
I will consider four expressions for the entropy,
which   use different approximations 
for 
the energy and particle number distributions.
A detailed descriptions of their properties 
are given in Sections 
\ref{section_Boltzmann},
\ref{section_Gibbs},
\ref{section_canonical},
and
\ref{section_grand_canonical},
but a brief overview is given here.

\begin{enumerate}
\item The simplest is the Boltzmann entropy,
which assumes that the number of particles 
is known exactly,
and the energy distribution is a delta function.

\item
Next comes the Gibbs entropy,
which also assumes that the number of particles 
is known exactly,
but that all states with energies below the energy of the system 
are counted.
This is not the correct energy distribution,
but it does provide a nonzero width.
It gives  the correct energy-temperature relationship
for a density of states that increases with energy.
If the density of states decreases with energy,
the predicted energy distribution is  incorrect.

\item
The canonical entropy
again assumes that the  number of particles 
is known exactly,
but uses the canonical ensemble 
to calculate the width the energy distribution
(see Subsection \ref{justification_of_canonical}).
It gives the  correct energy dependence of the entropy
for both increasing and decreasing densities of states.

\item
The grand canonical entropy
uses the grand canonical ensemble 
to calculate the entropy of both the energy and particle number 
distributions.  
It gives expressions for the entropy that are completely satisfactory.
A  striking feature 
is that the grand canonical entropy
of the classical ideal gas 
is exactly extensive --
without requiring Stirling's approximation.

\end{enumerate}

In  section~\ref{Macroscopic_probabilities},
% describes  the statistical basis 
%for the discussion of entropy.
% In subsequent sections,
I begin the discussion
 of the relevant statistical mechanics
 with the calculation of the probability 
distribution for energies, volumes, and number of particles 
for a large number of systems that might, or might not, interact.
I then discuss the Boltzmann and Gibbs entropies.
I  derive the canonical entropy.
Finally,
I  derive the grand canonical entropy,
which satisfies all thermodynamic conditions
I summarize the findings
in section \ref{section_summary}.

\section{Macroscopic probabilities}
\label{Macroscopic_probabilities}

The basic problem 
of thermodynamics 
is to predict the equilibrium
values of the extensive variables
 after the release of a constraint between systems\cite{Callen_1960,RHS_book}.
The solution to this problem 
in statistical mechanics
  does not require any assumptions about the proper definition of entropy.

 Consider a collection   of 
 $M \ge 2$ macroscopic systems,
 which 
 include all systems that
 might, or might not,
 exchange energy, volume, or particles.
 Denote the  phase space for the $j$-th  system 
by 
$\{p_j,q_j\}$,
where 
(in three dimensions)
$p_j$ represents $3N_j$ momentum variables,
and 
$q_j$ represents $3N_j$ configuration variables.
Making the usual assumption 
that interactions between systems are sufficiently short-ranged
that they may be neglected\cite{RHS_continuous},
 the total Hamiltonian 
 of the collection of systems
  can be written as 
a sum of contributions 
from each system.
\begin{equation}\label{H total 1}
H_T
=
\sum_{j=1}^M
H_j (p_j, q_j)
\end{equation}
The energy, volume, and particle number 
of system $j$
 are  denoted as
$E_j$, $V_j$, and  $N_j$\cite{footnote_1_particle_types},
and are subject to the conditions
on the sums,
\begin{equation}\label{sums}
\sum_{j=1}^{M} E_j   = E_T  , \,
\sum_{j=1}^{M} V_j   = V_T  , \,
\sum_{j=1}^{M} N_j   = N_T  ,
\end{equation}
where $E_T$, $V_T,$ and $N_T$ 
are constants.
The systems do not overlap each other.  
Naturally,
 only $3(N-1)$ 
 of the variables are independent.

I am not restricting the range of the interactions 
within any of the $M$ systems.
I am also not assuming homogeneity, 
%of the system,
so I do not, in general,
expect extensivity\cite{extensivity}.
For example,
the systems might be inclosed by adsorbing walls.
On the other hand,
I will
%, however,
use the classical ideal gas,
which is homogeneous and 
expected to be extensive,
as a simple example.

Since
I am  concerned with
 macroscopic experiments,
I assume that 
 no measurements
are made that might identify 
individual particles,
whether or not they are formally  indistinguishable\cite{RHS_distinguishability}.
Therefore, there are
$N_T!/ \left( \prod_{j=1}^{M}  N_j!\right)$
different permutations for assigning particles to systems,
and all permutations are taken to be equally probable.

The probability distribution
 for the macroscopic observables 
 in equilibrium 
can then be written as 
\begin{eqnarray}\label{W 1}
W \left( 
\{ E_j, V_j,N_j \}
%,t
 \right)
&=&
\frac{1}{\Omega_T}
\left(
\frac{
N_T!
}{
\prod_j N_j!
}
\right)  
  \nonumber  \\
&&
\times
\int  dp  \int   dq  
\prod_{j=1}^M
\delta( E_j - H_j )   ,
\end{eqnarray}
The constraint that 
the $N_k$ particles in
system $k$
are restricted to a volume 
$V_k$
is implicit in 
Eq.~(\ref{W 1}),
and the walls containing the system may have any desired properties.
$\Omega_T$
is a constant,
which is determined
%in principle 
 by summing or integrating over
all values of energy, volume, and particle number
that are consistent with 
the values of
$E_T$, $V_T$, and $N_T$
 in Eq.~(\ref{sums}).
 The value of the constant 
$\Omega_T$
does not affect the rest of the argument.

Eq.~(\ref{W 1})
 can also be written as 
\begin{equation}\label{W_2}
W(
 \{E_j, V_j, N_j 
\}
  ) 
=
\frac{1 }{  \Omega_T  }      
\prod_{j=1}^M
\Omega_j ( E_j, V_j, N_j )   ,
\end{equation}
where
\begin{equation}\label{Omega j 1}
\Omega_j
=
\frac{1}{h^{3N_j} N_j!}
\int_{-\infty}^{\infty}   d^{3N}p_j  \int   d^{3N}q_j   \,
\delta( E_j - H_j )   ,
\end{equation}
and
$\Omega_T$
is a normalization constant.
The factor of $1/h^{3N_j}$,
where $h$ is Planck's constant,
 is included  for
 agreement 
with the classical  limit 
of  quantum statistical mechanics\cite{RHS_book}.

For the classical ideal gas,
which I will use as an example for each proposed entropy,
\begin{equation}\label{eq_CIG_1}
\Omega_{CIG}(E,V,N)
=
\frac{V^N}{h^{3N} N!}
\frac{3N \pi^{3N/2} }{ (3N/2)!}
m
(2mE)^{3N/2-1}  .
\end{equation}
I have omitted the subscripts 
to make Eq.~(\ref{eq_CIG_1})
more compact.
The notation 
$ (3N/2)!$
is a convenient shorthand for
the more proper Gamma function.

\section{The Boltzmann entropy, $S_B$}
\label{section_Boltzmann}

Consider $M\ge 2$ systems 
%(which includes all systems that might interact)
with Hamiltonians 
$H_j(p_j,q_j)$.
The systems are originally isolated,
but individual constraints may be removed or imposed,
allowing 
 the possibility of  exchanging
energy, particles, or volume.
The number $M$ is intended to be quite large,
since all systems that might interact are included.
The magnitude of the energy involved in such potential interactions 
between systems
is regarded as negligible.
%
%Only the numbers of particles in each system is measured;
%individual particles are not identified,
%even if they are different,
%as is the case for colloids\cite{RHS_4}.
%
The probability distribution for the extensive thermodynamic variables,
energy ($E_j$),
volume ($V_j$),
and
number of particles ($N_j$),
is given by the expression 
in Eqs.~(\ref{W_2}) and (\ref{Omega j 1}).
The logarithm of Eq.~(\ref{W_2}) (plus an arbitrary constant, $C$)
gives the Boltzmann entropy of the $M$ systems.
\begin{equation}
S_T ( \{E_j, V_j, N_j \}  ) 
=
\sum_{j=1}^M
S_j ( E_j, V_j, N_j )
-  k_B   \ln    \Omega_T 
 +  C,
\end{equation}
where
\begin{equation}\label{S j = kB ln Omega j}
S_j  ( E_j, V_j, N_j  )
=
k_B  \ln  \Omega_j ( E_j, V_j, N_j  )  
\end{equation}
Using Eq.~(\ref{Omega j 1}),
\begin{equation}\label{S j = kB ln Omega j explicit}
S_j         % ( E, V, N  )
=
k_B  \ln 
\left[
\frac{1}{h^{3N_j} N_j!}
\int%_{-\infty}^{\infty} 
  d^{3N_j}p  \int   d^{3N_j}q   \,
\delta( E_j - H_j(p_j,q_j) )  
\right]  .
\end{equation}
Since the total Boltzmann entropy  is the logarithm
of 
the probability  
$W(\{E_j,V_j,N_j\})$,
maximizing the  Boltzmann entropy 
is equivalent to finding the mode of the probability distribution. 
This is not the same as finding the mean 
of the probability distribution,
but the difference
between the mean and the mode
is usually of order $1/N$.
As mentioned in the Introduction,
I will take even such small differences 
seriously.
% in the following sections.

For the classical ideal gas,
\begin{equation}
S_B(E,V,N)
=
k_B
\ln
\left(
\frac{V^N}{h^{3N} N!}
\frac{3N \pi^{3N/2} }{ (3N/2)!}
m
(2mE)^{3N/2-1}
\right)     ,
\end{equation}
or
\begin{eqnarray}\label{CIG_explicit}
S_B(E,V,N)
&=&
k_B
\left[
\ln
\left(
\frac{ E^{3N/2-1} }{ (3N/2)!}
\right)
\right.       \nonumber  \\
&&
\left.
+
\ln
\left(
\frac{V^N}{ N!}
\right)
+
\ln X^N
\right]  ,
\end{eqnarray}
where $X$ is a constant.  % given by
I will avoid using Stirling's approximation 
throughout the paper
to make all approximations explicit.

\subsection{Strengths  of the Boltzmann entropy}
\label{subsection_strengths_of_Boltzmann}

If any constraint between any two systems is released,
the probability distribution of the corresponding variable
is given by $W$ in Eq.~(\ref{W_2}).
Since the 
 Boltzmann entropy
 is proportional to the logarithm of $W$,
 it correctly predicts the mode
of the probability distribution.
Whenever the peak is narrow,
the mode is a very good estimate for the mean
since the relative difference is of the order of $1/N$,
while the fluctuations are of order $1/\sqrt{N}$.

\subsection{Weaknesses of the Boltzmann entropy}
\label{subsection_weaknesses_of_Boltzmann}

%The Boltzmann expression for the thermodynamic entropy is not perfect.

\begin{itemize}
\item 
$\Omega(E,V,N)$ 
has units of inverse energy,
so that it is not proper to take its logarithm.
This issue can  easily be resolved by multiplying 
$\Omega$ 
by a constant with units of energy before taking the logarithm in 
Eq.~(\ref{S j = kB ln Omega j explicit}).
This adds an arbitrary constant to the individual entropies,
but since it has no effect on any thermodynamic prediction,
it is not a serious problem. % for practical applications.

\item
The Boltzmann entropy can be shown 
\textit{not} to be 
adiabatically invariant\cite{Campisi_SHPMP_2005,%
Campisi_Kobe_2010,%
Campisi_SHPMP_2015%
}.
Campisi provides the following definition:
``A function  $I(E,V)$ is named an adiabatic invariant if, in the limit of very
slow variation of $V(t)$ namely as $t \rightarrow \infty$,
$I(E(t),V(t)) \rightarrow \textrm{const}$.''
The  violation of adiabatic invariance 
for the Boltzmann entropy
results in a relative error of the order of 
$1/N$.
The error is very small,
%(or even measurable for a macroscopic system),
but it is a weakness in the theory.
It is 
related to the use of the mode instead of the average.

\item
The assumption of a microcanonical ensemble
(that the probability distribution of the energy is a delta function)
is incorrect.
Since the width of the energy distribution is typically narrow,
of order $1/\sqrt{N}$, where $N$ is the number of particles,
the approximation is generally regarded as reasonable. 
It will turn out that this approximation is responsible 
for some of the weaknesses in the theory
and most of the disagreements.

\item
The number of particles $N$
is discrete.
The relative distance between
 individual points is very small,
 but nevertheless the entropy is not a continuous,
 differentiable function of $N$ 
 as assumed in thermodynamics.
 This discreteness  is usually simply ignored.

\item
The assumption that 
the number of particles  in our system
is known  \emph{exactly}
is never true for a macroscopic system.
The width of the distribution of  values of $N$ is  
much larger than the separation of points,
and $\langle N \rangle$ 
should be  used instead of $N$.

\item
The maximum of the Boltzmann entropy corresponds 
to the mode of the probability distribution --
not the mean.
This leads to small differences
 of order $1/N$.
For example, a partial derivative of the 
Boltzmann entropy with respect to energy
gives
\begin{equation}
U = (3N/2-1)k_B T  ,
\end{equation}
instead of 
\begin{equation}
U = (3N/2)k_B T  
\end{equation}
This error is  unmeasurable for macroscopic 
systems\cite{SW_2016_Physica,unmeasurable_error},
but it is a weakness of the theory.
It is the basis of
the argument against the validity
of the Boltzmann entropy.

\item
Although the classical ideal gas is composed 
of $N$ particles
that do not interact with each other,
the Boltzmann entropy is not exactly extensive
(see Eq.~(\ref{CIG_explicit})).     
We are  used to this lack of extensivity for finite $N$,
but it is incorrect.

\item
In the case of a first-order phase transition,
%the width of 
the energy distribution is not narrow,
and the  microcanonical ensemble
is not justified.
%This is confirmed by the fact that the
At a first-order transition,
a plot of
 the Boltzmann entropy 
 against energy
typically has a region of positive curvature,
although a well-known thermodynamic requirement for stability
states that the curvature must be negative\cite{Callen_1960,RHS_book}.
\begin{equation}
\pdbs{^2S}{E^2}{V,N} < 0
\end{equation}
This problem is a serious flaw in the theory.
I will discuss it in Sections \ref{section_canonical} and \ref{section_grand_canonical}.

\end{itemize}

\subsection{The 
approximation of
%source of the weaknesses in
 the Boltzmann entropy}

The Boltzmann entropy has assumed that the individual systems
have a microcanonical energy distribution (i.e.: a Dirac delta function in the energy).
This is a reasonable assumption, since the true distribution is known 
to be very narrow -- the relative width is of  order  $1/\sqrt{N}$.
However, 
a better approximation would be given by the canonical distribution,
which leads to the canonical entropy,
discussed in Section \ref{section_canonical_finite}.
Similarly,
the Boltzmann entropy
and the canonical entropy
both assume that the exact number of particles is known.
This is also a reasonable approximation,
but not really correct.
In Section \ref{section_grand_canonical},
the approximation of the distribution will again
be improved 
by using the grand canonical ensemble.

The three forms of the entropy,
Boltzmann, Canonical, and Grand Canonical,
are then based of successive improvements 
on the underlying probability distribution 
of the individual system.
It will become clear that the corresponding forms 
of the entropy are also successive improvements.

The Gibbs entropy,
introduced in the next section,
is an exception.
It is not closely related
to the energy distribution of the individual system,
except in the case of a
density of states
that is a monotonically increasing with energy.
For a decreasing density of states,
the distribution of states contributing to 
the Gibbs entropy
bears no resemblance to the 
probability distribution.

\section{The Gibbs entropy, $S_G$}
\label{section_Gibbs}

The Gibbs (or volume) entropy 
is defined by an integral over all energies 
less than the energy of the system\cite{Gibbs,Hertz_1910}.
It has the form
\begin{equation}\label{def_Gibbs}
S_G
=
k_B \ln \left[ \int_0^E \Omega(E',V,N) dE' \right]
\end{equation}

\subsection{Strengths of the Gibbs entropy}
\label{subsection_strengths_of_Gibbs}

\begin{itemize}
\item
The integral in the definition of the Gibbs entropy
in Eq.~(\ref{def_Gibbs})
is dimensionless,
so there is no problem in taking its logarithm.

\item
The Gibbs entropy can be shown 
 to be 
adiabatically invariant\cite{Campisi_SHPMP_2005,%
Campisi_Kobe_2010,%
Campisi_SHPMP_2015%
}.

\item 
For the Gibbs entropy  
 of classical systems with a montonically increasing density of states,
 the predicted energy is exactly 
correct\cite{DH_Physica_A_2006,%   {Phase transitions in small systems: Microcanonical vs. canonical ensembles
Campisi_Kobe_2010,%
DH_NatPhys_2014,%  Consistent thermostatistics forbids negative absolute temperatures
DH_reply_to_FW,%
HHD_2014,%   S. Hilbert and P. H\"anggi and J. Dunkel
Campisi_SHPMP_2015,%  Construction of microcanonical entropy on thermodynamic pillars
HHD_2016%   P. H\"anggi and  S. Hilbert and J. Dunkel
},
although this is not true of quantum systems\cite{RHS_continuous,Matty_comparison}.  
%
%For example,
%the Gibbs entropy predicts that the energy of the ideal gas is 
%$U =(3/2) N k_B T$.

\end{itemize}

\subsection{Weaknesses of the Gibbs entropy}
\label{subsection_weaknesses_of_Gibbs}

%The Gibbs expression for the thermodynamic entropy is  also not perfect.

\begin{itemize}

\item
The assumption of a microcanonical ensemble
(that the probability distribution of the energy is a delta function)
is  incorrect.

\item
The assumption that  
the  number of particles are in our system
is known \emph{exactly}
is  incorrect.

\item
The number of particles $N$
is  discrete,
and should be replaced by the continuous variable 
$\langle N \rangle$.

\item
Although the classical ideal gas is composed 
of $N$ particles
that do not interact with each other,
the Gibbs entropy is not exactly proportional
to $N$.
 This lack of extensivity of the Gibbs entropy 
is essentially the same as for the Boltzmann entropy.

\item
The Gibbs entropy also violates the thermodynamic inequality,
\begin{equation}
\pdbs{^2S}{E^2}{V,N} < 0  ,
\end{equation}
at a first-order transition.

\item
For a non-monotonic density of states,
the Gibbs entropy gives counter-intuitive results.
Consider two homogeneous systems 
with the same composition
and  a decreasing density of states
for the energies of interest
(for example: independent spins in a field).
Let them have the same energy per particle,
but let one be twice as large as the other.
 There will be no net transfer of energy
if the two systems are put in thermal contact.
The Boltzmann temperature will be the same for both systems,
as expected.
However,
the Gibbs temperature of the larger system 
will be higher.

\item
Because larger systems have higher Gibbs temperature,
it is impossible to construct a thermometer 
that measures the Gibbs temperature 
in an energy range with a  decreasing density of states.

\end{itemize}

The weaknesses of both the Boltzmann and the Gibbs entropies
can be avoided by using the
canonical and 
grand canonical ensembles,
instead of the microcanonical ensemble.

\section{The canonical entropy, $S_C$}
\label{section_canonical}

I have chosen to express the thermodynamic results in terms 
of Massieu functions\cite{Callen_1960},
because they do not require the inversion of the 
fundamental relation
$S=S(U,V,N)$.
The inversion 
to find   $U=U(S,V,N)$
is   unnecessary
and is not valid for systems with a non-monotonic
density of states.
%This is followed by the definition and analysis  of the canonical and grand canonical entropies.

\subsection{The general derivation of the canonical entropy}

Define a dimensionless entropy as
\begin{equation}
\tilde{S} = \frac{ S }{ k_B }
\end{equation}
Since 
\begin{equation}
dU = T dS - P dV + \mu dN
\end{equation}
and the inverse temperature is $\beta = 1/ k_B T$,
we also have
\begin{equation}\label{dS_1}
d \tilde{S} = \beta \, dU + \beta P dV - \beta \mu \, dN,
\end{equation}
where 
$P$ is the pressure,  $V$ is the volume,
$\mu$ is the chemical potential, and $N$ is the number of particles.
From Eq.~(\ref{dS_1}),
\begin{equation}\label{beta_dS_dU}          
\beta
=
\left(
\frac{ \partial \tilde{S} }{ \partial U }   
\right)_{V,N}  .
\end{equation}

The Legendre transform (Massieu function)
of $\tilde{S}$ 
with respect to $\beta$
is given by
\begin{equation}\label{S[b]=S-bU_1}                             % s[b]= - b F
\tilde{S}[\beta] = \tilde{S} - \beta U = - \beta \left( U - TS \right) = - \beta F ,
\end{equation}
so that 
\begin{equation}\label{St b = ln Z}
\tilde{S}[\beta] =\ln Z ( \beta, V,N )   .        % = - \beta F(T,V,N)  ,
\end{equation}
The differential  
of the Massieu function $\tilde{S}[\beta]$
is 
\begin{equation}
d \tilde{S}[\beta] = - U d \beta + \beta P dV - \beta \mu dN .
\end{equation}
This immediately gives 
\begin{equation}\label{d tilde S / dbeta = - U}
\left(
\frac{ \partial \tilde{S}[ \beta ] }{ \partial \beta  }   
\right)_{V,N}  =
- U    .
\end{equation}
%where the last equality is a well-known thermodynamic identity\cite{Callen_1985,RHS_book}.

To obtain $\tilde{S}$ from $\tilde{S}[\beta]$,
use 
\begin{equation}\label{Inverse_Masseu_C}
\tilde{S} = \tilde{S}[\beta] + \beta U   ,
\end{equation}
and substitute  $\beta = \beta(U)$.

\subsection{The justification of the canonical entropy}
\label{justification_of_canonical}

The use of the canonical ensemble 
to calculate the entropy of a general system 
requires some discussion.
That the probability distribution of the energy 
is not a delta function has been proven\cite{RHS_continuous}.
If the system of interest has been in contact with a much larger system
it is clear that the canonical ensemble is appropriate.
However,
if the system of interest  has instead been in contact with a system
that is the same size or even smaller,
it is known that the distribution is narrower than the canonical distribution\cite{Khinchin}.
Nevertheless,
the canonical entropy is appropriate
for calculating the thermodynamic entropy.

Consider three macroscopic  systems labeled 
$A$, $B$, and $C$.
Let systems 
$A$ and $B$   
be constructed to be exactly the same,
and in particular 
to be equal in size.
Let system $C$ be much larger than $A$ and $B$.

Suppose all three systems are  in thermal contact 
and have come to equilibrium.
The entropies of systems $A$ and $B$ are then equal,
and given by the canonical form discussed in the previous subsection.

Now remove system $C$ from thermal contact with 
the other two systems.
This is obviously a reversible process.
$A$ or $B$
are still in equilibrium with each other at the same temperature
as before $C$ was removed.
For consistency,
the entropies must be unchanged.
If the entropy were to decrease,
it would be a violation of the second law of thermodynamics.
If the entropy were to increase upon separation,
putting system $C$
back into thermal contact with systems $A$ and $B$
would decrease the entropy,
which would also violate the second law.
The only possibility consistent with the second law 
is that the entropy is unchanged.
Therefore,
the canonical entropy is properly defined 
for all systems,
regardless of their history or their size.

\subsection{The canonical entropy of the ideal gas}

Again we consider the special case of the 
classical ideal gas.
Start with the canonical  partition function.
\begin{eqnarray}\label{Z_CIG_1}
Z
&=&
\frac{1}{h^{3N} N!}
\int_{-\infty}^{\infty}   d^{3N}p  \int   d^{3N}q \,
\exp[ - \beta H ]    		\nonumber \\
&=&
\frac{1}{h^{3N} N!}
V^N
\int_{-\infty}^{\infty}   d^{3N}p  \,
\exp[ - \beta \sum_{j=1}^{3N}p^2_i / 2 m ]   \nonumber \\
&=&
\frac{1}{ N!}
V^N
\left(
\frac{2 \pi m }{ \beta h^2}
\right)^{3N/2}
\end{eqnarray}
Eq.~(\ref{St b = ln Z}),
%\begin{equation}\label{St b = ln Z 2_1}
%\tilde{S}[\beta] =\ln Z ( \beta, V,N )   .        % = - \beta F(T,V,N)  ,
%\end{equation}
%
\begin{equation}\label{St b = ln Z CIG_1}
\tilde{S}[\beta] 
=
\ln \left[
  \frac{1}{ N!}
V^N
\left(
\frac{2 \pi m }{ \beta h^2}
\right)^{3N/2}
\right]      ,
\end{equation}
and 
Eq.~(\ref{d tilde S / dbeta = - U}),
\begin{equation}
- U = \pdbs{ \tilde{S}[\beta] } {\beta} {V,N}   ,
\end{equation}
give the equation for $U$.
\begin{equation}\label{U_ideal_1}
U 
= 
-
\pd{}{\beta}
\ln 
\left[
  \frac{1}{ N!}
\left(
\frac{V}{h^3} 
\left(
\frac{2 \pi m }{ \beta}
 \right)^{3/2}
\right)^N
\right]
\end{equation}
Evaluating 
Eq.~(\ref{U_ideal_1})
gives
\begin{equation}\label{CIG_U_1}
 U 
= 
(3/2) N(1/\beta)   ,
\end{equation}
or,
\begin{equation}
U 
= 
\frac{3}{2} N k_B T   .
\end{equation}
This equation is exact,
while the usual Boltzmann entropy 
gives an error of order $1/N$.

To obtain $\tilde{S}$ from $\tilde{S}[\beta]$,
use 
\begin{equation}\label{Inverse_Masseu_2_1}
\tilde{S}
=
 \tilde{S}[\beta(U)] + \beta(U) U   .
\end{equation}
Since $\beta = 3 N / 2 U$ from Eq.~(\ref{CIG_U_1}),
\begin{equation}\label{St b = ln Z CIG_2}
\tilde{S}
=
\ln 
\left[
  \frac{1}{ N!}
\left(
\frac{V}{h^3} 
\left(
\frac{4 U \pi m }{ 3 N}
 \right)^{3/2}
\right)^N
\right]
+
\frac{3N}{2}
\end{equation}

Writing this out:
\begin{eqnarray}\label{St b = ln Z CIG_3}
S_C
&=&
k_B 
\left[
N
\left(
\frac{3 }{ 2}
 \right)
 \ln
 \left( \frac{U}{N}  \right)
  \right.
  +
 \ln \left( \frac{V^N }{ N! } \right)  \nonumber \\
 &&
 \left.
+
N
\left(
\frac{3 }{ 2}
 \right)
\ln
\left(
\frac{4  \pi m }{ 3  h^2}
 \right)
 +
\frac{3N}{2} 
\right]
\end{eqnarray}
Note that Eq.~(\ref{St b = ln Z CIG_3})
uses $U=\langle E \rangle$.
The factorial  $(3N/2)!$,
which comes from the surface area of a $3N$-dimensional
sphere in momentum space,
appears in the Boltzmann entropy,
but not  in the canonical entropy.
Only the term involving the volume $V$ involves a factorial
 ($1/N!$).
This comes from treating $N$ as a discrete variable,
which will be corrected by using 
the grand canonical ensemble
in Section \ref{section_grand_canonical}.

The problem of misrepresenting 
first-order phase transitions 
is also solved by using the 
canonical entropy\cite{Matty_comparison,Griffin_2017},
as shown in the next section.

\subsection{First-order phase transitions in the Canonical entropy}
\label{first_order_canonical}

The density of states for 
a system in which 
a first-order transition
is taking place
is characterized by a region of energies 
for which 
\begin{equation}
\frac{ \partial^2 \ln \Omega (E) }{ \partial E^2 } > 0   .
\end{equation}
This is a useful way of identifying 
first-order transition numerically.
However,
if the entropy is defined by either $S_B$ or $S_G$,
then these expressions will also have a region of energies
with
\begin{equation}
\frac{ \partial^2  S(E) }{ \partial E^2 } > 0   .
\end{equation}
This is forbidden by a well-known thermodynamic stability condition\cite{Callen_1960,RHS_book}.

A simple generic model
 density of states
  can be constructed
   that has this property.
\begin{eqnarray}\label{First_order_DoS_1}
\ln \Omega_1(E) 
&=&
 A N \left( \frac{E}{N} \right)^{\alpha}    
 + B  N \exp \left( - \frac{ E_{N,0}^2 }{ 2 \sigma^2_N } \right)     \nonumber \\
&& - B  N \exp \left( - \frac{ ( E - E_{N,0} )^2 }{ 2 \sigma^2_N } \right)  
\end{eqnarray}
The center of the Gaussian term is taken to be 
$E_{N,0} = f  N \epsilon$,
and the width of the Gaussian is  
$\sigma_N = g E_{N,0} $.
If the parameter $B$ does not depend on $N$,
this model corresponds 
to a mean-field transition,
while if $B$ decreases as  $N$ increases,
it corresponds to a system with short-range interactions.
The behavior is qualitatively the same in both cases.
I will treat the case of $B$ being constant,
because it is the more stringent test of the method.

\begin{figure}[htbp]
\begin{center}
 \includegraphics[width=\columnwidth]{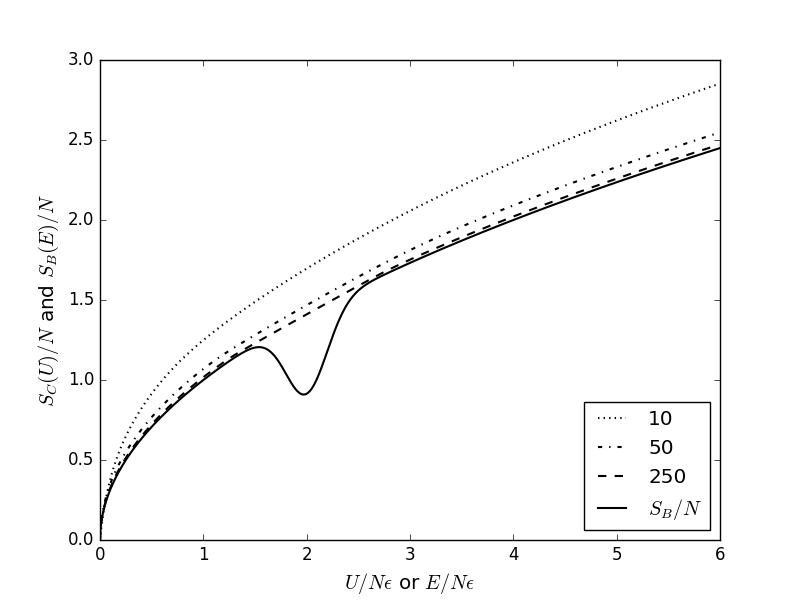}\\
  \caption{The top curves 
  (dotted, dot-dash, and dashed)
  indicate the 
  canonical entropy 
  $S_C/N$ 
  for $N=10,$  $50,$ and $250$.
  The solid curve shows $S_B$,
  which is the same in the model for all $N$.
The parameters are 
$\epsilon =1$,
$\alpha=0.5$,
$A=1$, $B=0.4$, $f=2$, and $g=0.1$,
but other values give similar results.
    }
  \label{SC_and_SB}
\end{center}
\end{figure}

Figure \ref{SC_and_SB}
shows 
$S_B(E)/N = k_B \ln \Omega(E)/N$
vs.\
$E/N \epsilon$.
The dip in the density of states 
and 
the region of positive curvature 
can be seen clearly.
Figure \ref{SC_and_SB}
 also shows
$S_C(U)/N$
  for $N=10,$  $50,$ and $250$,
  where $U=\langle E \rangle$.
  In all cases,
  the plot of $S_C$ shows 
  the negative curvature required for stability throughout.

%%   1

\subsection{Strengths of the Canonical entropy}
\label{subsection_strengths_of_canonical}

The canonical entropy for the thermodynamic entropy is superior to
than either the Boltzmann or the Gibbs entropies.

\begin{itemize}

\item
The canonical entropy is  adiabatically invariant. 

\item
The canonical entropy gives a thermodynamically 
correct description of 
first-order transitions.

\item 
The  canonical entropy has an 
energy term of
$(3/2)
 k_B N 
 \ln
 \left( U/N \right)$
 for the ideal gas,
 which is correct.
 As a consequence,
  the 
canonical entropy 
also gives the exact energy
for the classical ideal gas.
\begin{equation}
U = (3N/2)k_B T 
\end{equation}
\end{itemize}

\subsection{Weaknesses of the Canonical entropy}
\label{subsection_weaknesses_of_canonical}

%The canonical entropy  is still not perfect.

\begin{itemize}

\item
The canonical entropy
assumes that the value of $N$ is known exactly,
instead of using 
 a distribution of possible values of $N$.

\item
The number of particles $N$
is  discrete.
%The consequences of removing this approximation  will be discussed in section \ref{section_grand_canonical}.

%

\item
The assumption that the number of particles is known exactly is incorrect.

\item
The deviation from exact extensivity
(in the factor $1/N!$) is a weakness.
 This lack of extensivity of the canonical entropy  
differs from that of the Boltzmann entropy,
in that it does not affect the energy-dependent
term.

\end{itemize}

\section{The grand canonical entropy, $S_{GC}$}
\label{section_grand_canonical}

The grand canonical entropy satisfies all criteria 
required by thermodynamics.

\subsection{The  definition of the grand canonical Legendre transform}

For the grand canonical ensemble,
$S[\beta,(\beta \mu)]$
will be used.
I have put parentheses around the second variable 
to emphasize that the product of $\beta$ and $\mu$
is to be treated as a single variable.
To find the Legendre transform with respect to both
$\beta$ and $(\beta \mu)$
use the equation 
\begin{equation}\label{beta_mu_dS_dN}          
- (\beta \mu )
=
\left(
\frac{ \partial \tilde{S} }{ \partial N }   
\right)_{U,N}  .
\end{equation}
in addition to
Eq.~(\ref{beta_dS_dU}).

The Legendre transform (Massieu function)
of $\tilde{S}$ 
with respect to both $\beta$ and
$(\beta \mu)$
is given by
\begin{equation}\label{S[b]=S-bU_2}                             % s[b]= - b F
\tilde{S}[\beta,(\beta \mu)] = \tilde{S} - \beta U  + (\beta \mu) N,
\end{equation}
so that 
\begin{equation}\label{St b = ln Z_3}
\tilde{S}[\beta,(\beta \mu)] =\ln \mathcal{Z} ( \beta, V,  (\beta \mu) )   ,       % = - \beta F(T,V,N)  ,
\end{equation}
where $\mathcal{Z}$ is the grand canonical partition function.

The differential  
of the Massieu function $ \tilde{S}[\beta,(\beta \mu)]$
is 
\begin{equation}
d \tilde{S}[\beta,(\beta \mu)] = - U d \beta + \beta P dV + N d( \beta \mu) .
\end{equation}
This immediately gives
\begin{equation}\label{d tilde S / dbeta = - U gc}
\left(
\frac{ \partial\tilde{S}[\beta,(\beta \mu)]  }{ \partial \beta  }   
\right)_{V,(\beta \mu)}  =
- U   ,
\end{equation}
and
\begin{equation}\label{d tilde S / d(beta mu) = - U gc}
\left(
\frac{ \tilde{S}[\beta,(\beta \mu)] }{ \partial (\beta \mu) }   
\right)_{\beta,V}  =
N   .
\end{equation}

To obtain $\tilde{S}$ from $\tilde{S}[\beta,(\beta \mu)] $,
use 
\begin{equation}\label{Inverse_Masseu_B}
\tilde{S} = \tilde{S}[\beta,(\beta \mu)]  + \beta U  -( \beta \mu) N ,
\end{equation}
and replace the $\beta$ and $(\beta \mu)$ dependence 
by $U$ and $\langle N \rangle$.

\subsection{The grand canonical entropy of the ideal gas}

For the example of the classical ideal gas,
the
canonical partition function is given in 
Eq.~(\ref{Z_CIG_1}).
\begin{equation}\label{Z_CIG_3_2}
Z
=
\frac{1}{ N!}
V^N
\left(
\frac{2 \pi m }{ \beta h^2}
\right)^{3N/2}
\end{equation}
To obtain the grand canonical partition function,
multiply this by 
$\exp[ (\beta \mu) N]$
and sum over $N$.
\begin{equation}\label{Z_CIG_3_3}
\mathcal{Z}
=
\sum_{N=0}^{\infty}
\frac{1}{ N!}
%V^N
\left(
V
\left(
\frac{2 \pi m }{ \beta h^2}
\right)^{3/2}
\right)^N
\exp[ (\beta \mu) N]
\end{equation}
The series sums to an exponential.
\begin{equation}\label{Z_CIG_3_4}
\mathcal{Z}
=
\exp \left[
\left(
V
\left(
\frac{2 \pi m }{ \beta h^2}
\right)^{3/2}
\right)
\exp[ (\beta \mu) ]
\right]
\end{equation}

The Massieu function for the grand canonical ensemble is
\begin{equation}\label{St b = ln Z 2_2}
\tilde{S}[\beta,(\beta \mu)] =\ln \mathcal{Z} ( \beta, V,N )   .        % = - \beta F(T,V,N)  ,
\end{equation}
\begin{equation}\label{St b = ln Z CIG_4}
\tilde{S}[\beta,(\beta \mu)] 
=
\left(
V
\left(
\frac{2 \pi m }{ \beta h^2}
\right)^{3/2}
\right)
\exp[ (\beta \mu) ]
\end{equation}

The average number of particles is given by
\begin{equation}
 \langle N \rangle
 = \pdbs{ \tilde{S}[\beta,( \beta \mu)] } {(\beta \mu)} {\beta,V}
 = \tilde{S}[\beta,( \beta \mu)] .
\end{equation}
This equation can be used to solve for 
 $(\beta \mu)$.
 \begin{equation}
 \langle N \rangle
=
\left(
V
\left(
\frac{2 \pi m }{ \beta h^2}
\right)^{3/2}
\right)
\exp[ (\beta \mu) ]
\end{equation}
 \begin{equation}
 (\beta \mu) 
=
\ln
\left[
\frac{ \langle N \rangle}{V}
\left(
\frac{ \beta h^2}{2 \pi m }
\right)^{3/2}
\right]
\end{equation}

To find the energy, use
\begin{equation}
- U =  \pdbs{ \tilde{S}[\beta, \beta ] } {(\beta )} {V, ( \beta \mu)},
\end{equation}

\begin{equation}
- U = 
-\frac{3}{2}
\beta^{-5/2}
\left(
V
\left(
\frac{2 \pi m }{  h^2}
\right)^{3/2}
\right)
\exp[ (\beta \mu) ]   ,
\end{equation}
or,
using 
$  \langle N \rangle = \tilde{S}[\beta,( \beta \mu)] $
\begin{equation}
U = 
\frac{3}{2}
\beta^{-1}
 \langle N \rangle
\end{equation}

To obtain $\tilde{S}$ from $\tilde{S}[\beta,(\beta \mu)]$,
use the inverse Legendre transform.
\begin{equation}\label{Inverse_Masseu_2_2}
\tilde{S} 
=
 \tilde{S}[\beta(U, \langle N \rangle), (\beta \mu)] + \beta(U, \langle N \rangle) U  -(\beta \mu)  \langle N \rangle 
\end{equation}
\begin{equation}
\tilde{S} 
=
  \langle N \rangle 
 +
\frac{3 \langle N \rangle}{2}
  -
   \ln
\left[
\frac{ \langle N \rangle}{V}
\left(
\frac{ \beta h^2}{2 \pi m }
\right)^{3/2}
\right]  \langle N \rangle 
\end{equation}

Inserting $\beta(U, \langle N \rangle)$
and
using $S = k_B \tilde{S} $
gives the grand canonical entropy.
%for the classical ideal gas.
\begin{eqnarray}
S_{GC}
&=&
  \langle N \rangle  k_B
\left[
\frac{3}{2}
\ln
\left(
\frac{U}{\langle N \rangle} \right)
+
\ln
\left(
\frac{V}{\langle N \rangle}
\right) 
\right. \nonumber \\
 &&
 \left.
+
\ln
\left(
\frac{ 4 \pi m }{ 3 h^2}
\right)^{3/2}
+
\frac{5}{2}
\right] 
\end{eqnarray}

This expression for the entropy
of a classical ideal gas is exactly extensive\cite{extensivity},
and no use has been made of Stirling's approximation.

\subsection{Strengths of the grand canonical entropy}
\label{subsection_strengths_of_grand_canonical}

The grand canonical entropy retains the advantages of the canonical entropy,
but also has a correct description of the distribution of particles.
It provides a completely consistent  description 
of the properties of a thermodynamic system.

The grand canonical entropy for the 
classical ideal gas
is exactly extensive,
which is  expected
of a model in which there are no explicit interactions 
between particles.
%but which is not found for  $S_B$, $S_B$, or $S_C$.

%
\subsection{Weaknesses of the grand canonical entropy}
\label{subsection_weaknesses_of_grand_canonical}

None.

\section{Summary}
\label{section_summary}

I have discussed the properties of four different definitions
of the classical entropy from statistical mechanics:
the Boltzmann, the Gibbs, the canonical, and the grand canonical.
I have shown that the microcanonical assumption
(that the energy is known exactly)
is responsible for  weaknesses in the 
Boltzmann  entropy,
and that the 
grand canonical entropy satisfies
all thermodynamic requirements.

The original arguments against 
negative temperatures were 
actually aimed at the Boltzmann entropy.
%which predicted them.
In particular,
the violation of adiabatic invariance 
was regarded as being thermodynamically inconsistent,
even though the violation was very small.
The grand canonical entropy
%$S_{GC}$ 
does not suffer from any of such weakness.
%The grand canonical entropy 
It
gives consistent thermodynamics for 
all
 %both positive and negative 
 temperatures,
 effectively removing the original argument
 against negative temperatures.

The grand canonical entropy
has other advantages.

For models of independent particles,
the entropy is expected to be exactly extensive,
which is the case  for
the grand canonical entropy,
but not for the other three definitions.

First-order transitions were shown to be correctly represented 
by the canonical and grand canonical ensembles 
in Section \ref{first_order_canonical},
but  by neither the Boltzmann nor the Gibbs entropies.
The same was shown to be true 
for quantum systems    
%statistical mechanics
in Ref.~\cite{Matty_comparison}.

The Gibbs entropy has much the same behavior 
as the Boltzmann entropy
\textit{for a monotonically increasing density of states}.
The difference is only in the energy dependence of the entropy, 
which is correct
for the Gibbs entropy
(and $S_C$ and $S_{GC}$),
while the Boltzmann entropy has an error of order $1/N$.
The both the Boltzmann and the Gibbs entropy
% is  only defined on integer values of $N$,  and
have  $\ln(N) /N$
errors in the $N$-dependence.
%It  gives incorrect results  at first-order phase transitions.

The Gibbs entropy has serious problems 
\textit{for a  decreasing density of states},
for which it  is the only definition that fails to 
predict negative temperatures.
The Gibbs entropy fails to predict the correct equilibrium values of the energy
for a decreasing density of states.
Equality of the  Gibbs temperature for systems of different sizes
does not predict that there will be no net transfer of energy
if the two systems are put in thermal contact with each other.
The Boltzmann, canonical, and grand canonical  entropies
all predict the physical behavior correctly.

The question of whether negative temperatures 
are thermodynamically consistent
is most naturally discussed in quantum statistical mechanics.
Various models 
 were  investigated
in Ref.~\cite{RHS_continuous} and \cite{Matty_comparison},
included independent spins and the two-dimensional,
twelve-state Potts model. 
The  canonical entropy describes 
the behavior of these models
with negative temperatures,
and without the weaknesses that have sparked criticism 
of the Boltzmann entropy.

The Boltzmann entropy, $S_B$,
has received a great deal of criticism
%in the history of the controversy 
during the controversy 
concerning negative temperatures\cite{Campisi_SHPMP_2005,%
DH_Physica_A_2006,%   {Phase transitions in small systems: Microcanonical vs. canonical ensembles
Campisi_Kobe_2010,%
Romero-Rochin_2013,%
Sokolov_2014,%
DH_NatPhys_2014,%  Consistent thermostatistics forbids negative absolute temperatures
DH_reply_to_FW,%
HHD_2014,%   S. Hilbert and P. H\"anggi and J. Dunkel
Campisi_SHPMP_2015,%  Construction of microcanonical entropy on thermodynamic pillars
HHD_2016%   P. H\"anggi and  S. Hilbert and J. Dunkel
}.
However,
after recognizing that the grand canonical entropy %,
%$S_{GC}$,
is correct,
we can  see that the 
Boltzmann entropy
provides an excellent description of thermodynamics,
if
the missing factor of the energy in $\Omega(E,V,N)$ 
is ignored,
and Stirling's approximation is used
whenever factorials are encountered.
In other words,
the Boltzmann entropy is usually quite satisfactory,
if we agree to ignore %unmeasurable 
errors of the order of $1/N$.

The exception to the validity of the Boltzmann entropy
is a first-order phase transition.
In that case,
the energy distribution is broad
(violating an assumption made in the derivation 
of the Boltzmann entropy),
and the canonical or grand canonical entropy should used.
However,
a good approximation to
the true entropy can be  obtained by the well-known double-tangent construction.
This will fail to show the finite-size rounding,
but is otherwise satisfactory.

These results should settle the controversy 
in favor of the thermodynamic validity of negative temperatures.
The Gibbs entropy is useful only for models with increasing densities of states.
The correct entropy is given by the grand canonical formulation,
but the Boltzmann entropy is generally satisfactory.

\section*{Acknowledgement}

I would like to thank 
Jian-Sheng Wang and 
Johannes Zierenberg 
for useful comments.
I would also like the thank 
J. B.  Kadane and Roberta Klatzky for many helpful discussions.
This research did not receive any specific grant from funding agencies 
in the public, commercial, or
not-for-profit sectors.

\makeatletter
\renewcommand\@biblabel[1]{#1. }
\makeatother

\bibliography{Entropy_citations_3}

\begin{thebibliography}{44}
\bibitem [{Gib()}]{Gibbs_entropy_2}%
  \BibitemOpen
  \href@noop {} {}\bibinfo {note} {I am using the term ``Gibbs entropy'' to
  refer to the definition of the entropy in terms of the logarithm of the
  volume of phase space with energy less than a given energy. It is not to be
  confused with another definition of entropy due to Gibbs in terms of an
  integral of $\int \rho \ln \rho$, where $\rho$ is the probability of a
  microscopic state.}\BibitemShut {Stop}%
\bibitem [{Bol()}]{Boltzmann_classical_entropy}%
  \BibitemOpen
  \href@noop {} {}\bibinfo {note} {Boltzmann derived the expression for the
  entropy for classical systems under the assumption that equilibrium
  corresponded to the maximum of the probability for two systems in
  equilibrium.}\BibitemShut {Stop}%
\bibitem [{foo()}]{footnote_1_particle_types}%
  \BibitemOpen
  \href@noop {} {}\bibinfo {note} {I will only write the equations for a single
  type of particle. The generalization to a variety of particles is trivial,
  but requires indices that might obscure the essential argument.}\BibitemShut
  {Stop}%
\bibitem [{ext()}]{extensivity}%
  \BibitemOpen
  \href@noop {} {}\bibinfo {note} {I am distinguishing extensivity from
  additivity. The entropy of a system is extensive when $\lambda S(U,V,N) =
  S(\lambda U,\lambda V, \lambda N)$. The entropies of two systems are additive
  when $S_{A,B} = S_A+S_B$.}\BibitemShut {Stop}%
\end{thebibliography}

\end{document}